\documentclass{ws-procs10x7}
\usepackage{balance}

\newcolumntype{d}[1]{D{.}{.}{#1}}

\def\Journal#1#2#3#4{{\it #1} {\bf #2}, #3 (#4)}

\makeindex
\begin{document}

\title{COLLECTIVE EFFECTS IN A MEDIUM AND A MODEL OF COMPOUND FLOW
IN RELATIVISTIC HEAVY ION COLLISIONS}

\author{V. A. OKOROKOV}

\address{Department of Physics, Moscow Engineering Physics Institute (State
University),\\Kashirskoe Shosse 31, 115409 Moscow, Russian
Federation\\E-mail: Okorokov@bnl.gov;~VAOkorokov@mephi.ru}


\twocolumn[\maketitle\abstract{A new analytical approach is
presented for analysis of two-particle azimuthal correlations in
heavy ion collisions at relativistic energies. This approach
suggests that elliptic flow measured by experiment has a compound
structure, namely, that it may come from superposition of several
components. General expressions have been derived for the
two-particle correlation function isolating the contribution due
to anisotropic flow. The model of compound flow takes into account
the number of jets per event, average multiplicity per jet,
dependence of jet yield on the orientation with respect to the
reaction plane, and independent "soft" particle production. These
analytic calculations provide the framework for a consistent
description of the elliptic flow measured via the single-particle
distribution with respect to the reaction plane, jet yield per
event, and the amplitude of flow-like modulation in the
two-particle distribution in the relative azimuthal angle.}
 ]

\section{Introduction}
The specific fundamental prediction of QCD most often highlighted
in the discussions of experimental program at RHIC (BNL) and LHC
(CERN) is that at high energy densities an ordinary hadronic
matter undergoes a phase transition to a new state of hot and
dense strongly interacting matter, the QGP, and reproduce the
early Universe conditions. A study of collective behavior in
heavy-ion collisions provides one of the most sensitive and
promising probes for investigation of possible QGP formation and
elucidating its properties.

Large collective flow measured in Au-Au collisions at the RHIC
provides direct evidence of strong pressure gradients in a hot and
dense medium created in a collision. Quenching of di-jets observed
in central Au-Au collisions at RHIC by studying the two-particle
azimuthal distributions at high transverse momenta
\cite{Adams-2004} is interpreted as final state effect and as an
evidence for extreme energy loss of partons propagating in the hot
deconfined matter \cite{Adams-2005}. At intermediate transverse
momenta, both effects are present in the azimuthal correlations of
hadron pairs and it is important to disentangle the contributions
of flow and jet components.

Thus azimuthal correlations of hadron pairs provide one of the
primary tools for the experimental investigation of collective
flow phenomena and identifying the products of hard jets. A new
analytical approach is presented for analysis of two-particle
azimuthal correlations in heavy ion collisions at relativistic
energies.

\section{Two-particle correlations: disentangling flow and jets}
In the following, we suggest a general framework for analysis of
the azimuthal correlations of hadron pairs in heavy-ion
experiments at high (RHIC and LHC) energies. Two-particle
distributions in the relative azimuthal angle of charged high
$p_T$ particles measured in p+p, d+Au, and Au+Au collisions at
RHIC exhibit a jet-like correlation characterized by the peaks at
$\Delta\phi=0$ (near-side jets) and at $\Delta\phi=\pi$
(opposite-side jets). In Au+Au collisions, the particle
correlations due to elliptic flow also contribute, resulting in a
cosine-like pattern. We suggest that elliptic flow measured by
experiment has a compound structure, namely, that it may come from
superposition of several components. We derive general expressions
for the two-particle correlation function isolating the
contribution due to anisotropic flow. We take into account the
number of jets per event, average multiplicity per jet, dependence
of jet yield on the orientation with respect to the reaction
plane, and independent "soft" particle production. The different
distributions are normalized by standard way, namely, all
single-particle distributions are normalized on total number of
particles per event and all type two-particle distributions are
normalized on total pair number per event below.

Anisotropic elliptic flow primarily manifests itself by the
presence of the second Fourier coefficient ($v_2$) in the
inclusive single particle distribution in the azimuthal angle
$\phi$ with respect to the reaction plane: $$2\pi\frac{\textstyle
dn^{s}}{\textstyle
d\phi}=N_{s}\left[1+2v^{s}_{2}\cos2\phi\right],$$ where $N_s$ is
the number of "soft" particles, or particles which have no other
azimuthal correlations besides those due to elliptic flow.
Elliptic flow also generates azimuthal anisotropy in the angle
difference $\Delta\phi$ of "soft" particle pairs:
\begin{equation}
2\pi\frac{\textstyle dn^{s}_{1,2}}{\textstyle
d\left(\Delta\phi\right)}=N^{p}_{s}
\left[1+2\left(v^{s}_{2}\right)^{2}\cos2\Delta\phi\right],\label{eq:soft}
\end{equation}
where  $N^{p}_{s}=N_{s}\left(N_{s}-1\right)$ is the number of
"soft" particle pairs. Suppose that for events where hard
scattering took place, the $N_{h}$ "hard" particles were produced
via fragmentation of high energy partons. These particles may also
be correlated with the impact parameter due to parton energy loss
and the azimuthal dependence of the path length, but in general
case their degree of correlation ($v^{h}_{2})$ will be different
than that of the "soft" particles. The single-particle azimuthal
distribution of "soft" and "hard" particles with respect to the
reaction plane is then:
\begin{equation}
2\pi\frac{\textstyle dn}{\textstyle
d\phi}=N\left[1+2\tilde{v_{2}}\cos2\phi\right],
\label{eq:softhard}
\end{equation}
where $N=N_{s}+N_{h}$ -- total multiplicity of particle satisfied
some cuts per event. Thus, experiments measure the "effective"
$\tilde{v_2}=(N_{s}v^{s}_{2}+N_{h}v^{h}_{2})/(N_{s}+N_{h})$.

Hard scattered partons fragment into a high energy cluster (jet)
of hadrons which are distributed in a cone of size
$\Delta\eta\Delta\phi\sim0.7$ in pseudorapidity and azimuth. Let
the two-particle azimuthal distribution within a single jet/dijet
be of some general form:
$$2\pi\frac{\textstyle dn^{h}_{1,2}}{\textstyle
d\left(\Delta\phi\right)}=\left<n^{h}\right>\left(\left<n^{h}\right>-1\right)
f\left(\Delta\phi\right),$$
where $\left<n^{h}\right>$ is the average multiplicity of
particles within a single jet/dijet, and
$f\left(\Delta\phi\right)$ is a function which describes the
intra-jet correlations of hadron pairs within a jet cone. In
elementary high-energy collisions, $f\left(\Delta\phi\right)$
closely resembles the Gaussian distribution. In heavy-ion
collisions, we are interested in measuring possible medium
modifications of $f\left(\Delta\phi\right)$. The function
$f\left(\Delta\phi\right)$ can be parameterized as following:
\begin{eqnarray}
f\left(\Delta\phi\right)=A_{N}\frac{\textstyle 1}{\textstyle
\sqrt{2\pi \sigma_{N}^{2}}}~\exp\left(-\frac{\textstyle
\left(\Delta\phi\right)^{2}}{\textstyle 2\sigma_{N}^{2}}\right)\nonumber \\[4pt]
 + A_{B}\frac{\textstyle 1}{\textstyle \sqrt{2\pi
\sigma_{B}^{2}}}~\exp\left(-\frac{\textstyle \left(\Delta\phi -
\pi \right)^{2}}{\textstyle 2\sigma_{B}^{2}}\right) + \varepsilon,
\label{eq:diff-gauss}
\end{eqnarray}
where the $A_{N},~\sigma_{N}$ -- normalization factor and width
for the near-side peak $\left(\Delta\phi=0\right)$,
$A_{B},~\sigma_{B}$ -- normalization factor and width for the
opposite-side peak $\left(\Delta\phi=\pi\right)$, and
$\varepsilon$ -- (unknown) factor which is corresponds to medium
modification of hadron jets. The problem of normalization of
two-particle distribution doesn't discussed in this paper in
detail. Experimental distributions are normalized on number of
trigger particles with high $p_{t}$ and on detector efficiency for
corresponding single track. There are more complete normalization
descriptions for experimental distributions and the numerical
estimations of corresponding parameters in the
Eq.(\ref{eq:diff-gauss}) for different colliding systems and
energies \cite{Adams-2004,Adler-2003}.

Suppose we have $N_{J}$ jets per event\footnote{that is $N_{J}$
acts of hard parton scattering per event}, the total number of
particles from jets is then $N_{h}=N_{J}\left<n^{h}\right>.$
Combining the intra-jet correlations with correlations with
respect to the impact parameter, the two-particle azimuthal
distribution for "hard" particles is then given by the following
equation:
\begin{eqnarray}
2\pi\frac{\textstyle dn^{h}_{1,2}}{\textstyle
d\left(\Delta\phi\right)}=N_{h}\left(N_{h}/N_{J}-1\right)f\left(\Delta\phi\right)\nonumber
\\[4pt] \hspace*{0.5 cm}+ N^{p}_{h}
\left[1+2\left(v^{h}_{2}\right)^{2}\cos2\Delta\phi\right],
\label{eq:hard}
\end{eqnarray}
where the first term describes the correlation of particles from
the same jet/dijet, and the second term corresponds to the
correlation of particles from different jets/dijets (pure elliptic
flow of "hard" particles),
$N^{p}_{h}=N_{h}^{2}\left(1-1/N_{J}\right)$ -- number of "hard"
particle pairs per event with the exception of pairs between
"hard" particles within single jet.

Thus the combined two-particle distribution of "soft" and "hard"
particles will be a superposition of the "soft" and "hard"
distributions and an additional cross term with standard
normalization:
\begin{eqnarray}
2\pi\frac{\textstyle dn_{1,2}}{\textstyle
d\left(\Delta\phi\right)}&=&N_{h}\left(N_{h}/N_{J}-1\right)f\left(\Delta\phi\right)\nonumber
\\[4pt] &&{} + C\left[1+2P\cos2\Delta\phi\right], \label{eq:softAndhard}
\end{eqnarray}
where the new parameters are
$C=N^{p}_{s}+N^{p}_{h}+2N_{s}N_{h},
PC=N^{p}_{s}\left(v^{s}_{2}\right)^{2}+N^{p}_{h}
\left(v^{h}_{2}\right)^{2}+2N_{s}N_{h}v_{2}^{s}v_{2}^{h}.$

As one can see from the expression above, the coefficient $P$ in
front of $\cos2\Delta\phi$ depends on five parameters, namely
$N_{J}, N_{s}, N_{h}, v_{2}^{s}$, and $v_{2}^{h}$ in non-trivial
way. This coefficient is the square of "generalized" flow in the
framework of this model of two-component flow. The total
multiplicity $\left(N_{s}+N_{h}\right)$ is measured
experimentally. Experiments also measure
$\tilde{v_2}=(N_{s}v^{s}_{2}+N_{h}v^{h}_{2})/(N_{s}+N_{h})$. More
information can be extracted from the experimental data using
conditional two-particle azimuthal correlations for which one of
the particles is detected under fixed directions with respect to
the reaction plane \cite{Bielcikova-2004}.

 We also investigate how
the two-particle azimuthal distributions change when the
orientation of one of the particles is restricted to a region
in-plane or out-of-plane with respect to the reaction plane. The
analytic calculations below provide the framework for a consistent
description of the elliptic flow measured via the single-particle
distribution with respect to the reaction plane, jet yield per
event, and the amplitude of flow-like modulation in the
two-particle distribution in the relative azimuthal angle. Let the
trigger particle be confined in the transverse plane to the
$-\pi/4 < \phi<\pi/4 $ ({\it 'in-plane'}~), and $\pi/4 < \phi<
3\pi/4$ ({\it 'out-of-plane'}~), respectively. Below we consider
the simplest ideal case, namely, the case of known reaction plane.
The number of in/out "soft" and "hard" particles is:
$$\left(N_{out}^{in}\right)_{s}=
\frac{\textstyle N_{s}}{\textstyle 2\pi}\left(\pi \pm
4v_{2}^{s}\right),$$
$$\left(N_{out}^{in}\right)_{h}=
\frac{\textstyle N_{h}}{\textstyle 2\pi}\left(\pi \pm
4v_{2}^{h}\right).$$

There are four terms to the total number of possible pairs for the
combined "soft+hard" in/out-plane two-particle distribution:
\begin{eqnarray}
2\pi\left(\frac{\textstyle dn_{1,2}}{\textstyle
d\left(\Delta\phi\right)}\right)^{in}_{out}&=&
\mbox{soft}^{in}_{out} \cdot \mbox{soft}\nonumber \\[4pt]&&{}+\mbox{soft}^{in}_{out}
\cdot \mbox{hard}\nonumber \\[4pt]&&{}+\mbox{hard}^{in}_{out} \cdot
\mbox{soft}\nonumber \\[4pt]&&{}+\mbox{hard}^{in}_{out} \cdot
\mbox{hard}.\label{eq:InOut-comb-1}
\end{eqnarray}

By adding up all four terms and combining, we get the following
general expression for two-particle "in/out-plane" distribution:
\begin{eqnarray}
2\pi\left(\frac{\textstyle dn_{1,2}}{\textstyle
d\left(\Delta\phi\right)}\right)^{in}_{out}=\left(N^{in}_{out}\right)_{h}
\left(\frac{\textstyle N_{h}}{\textstyle
N_{J}}-1\right)f\left(\Delta\phi\right)\nonumber
\\[4pt]+
C^{in}_{out}\left[1+2P^{in}_{out}\cos2\Delta\phi\right],\nonumber
\end{eqnarray}
\begin{eqnarray}
2\pi C^{in}_{out} \equiv \tilde{C}^{in}_{out}&=&\left(\pi \pm
4v_{2}^{s}\right)
\left(N^{p}_{s}+N_{s}N_{h}\right)\nonumber \\[4pt]&&{}+\left(\pi \pm
4v_{2}^{h}\right) \left(N_{s}N_{h}+N^{p}_{h}\right),\nonumber
\end{eqnarray}
\begin{eqnarray}
\tilde{C}^{in}_{out}P^{in}_{out}&=&N^{p}_{s}v_{2}^{s}\left(\pi
v_{2}^{s}\pm 2\right)+N^{p}_{h}v_{2}^{h}
\left(\pi v_{2}^{h} \pm 2\right)\nonumber \\[4pt]&&{}+ 2N_{s}N_{h}\left(\pi
v_{2}^{s}v_{2}^{h}\pm v_{2}^{s} \pm v_{2}^{h}\right).
\label{eq:InOut-softAndhard}
\end{eqnarray}

By performing the combined fit of Eqs.(\ref{eq:softhard}),
(\ref{eq:softAndhard}) and (\ref{eq:InOut-softAndhard}) to the
experimental data, constraining the integrals, and using the fact
that flow cancels out at $\phi,\Delta\phi=\pi/2$, it may be
possible to unambiguously determine parameters $N_{J}, N_{s},
N_{h}, v_{2}^{s}, v_{2}^{h}$ and study the medium modification of
the two-particle distribution within a jet by the following way.

Thus jets are sensitive to the collective flow field in the
collision region. The reason of the jet asymmetry can be both
collective flow of medium (flow of "soft" particles) and flow of
"hard" particles (i.e. jet correlations with respect to event
plane - jet flow). Also the "hard" flow component can influences
on the structure of away-side peak in hadron distribution for RHIC
data \cite{Wang-2004}. In particular the jet flow can provides
some asymmetry of Max cone or ring of Cherenkov gluons
\cite{Dremin-2006}.

\section*{Summary}
We derived general expressions for the two-particle azimuthal
distributions isolating the contributions due to anisotropic flow
and jets. We took into account the number of jets per event,
average multiplicity per jet, dependence of jet yield on the
orientation with respect to the reaction plane, and independent
"soft" particle production. In this model the two-particle
distribution in the relative azimuthal angle $\Delta \phi$ in
general case contains both "soft" and "hard" flow parameters in
non-trivial way. Analytical solutions are derived for two-particle
azimuthal distributions with respect to reaction plane in general
case for the first time.

The study of jet observables and jet-like particle correlations in
heavy ion collisions is still at the beginning. Experimental
methods for studying compound flow in relativistic heavy ion
collisions are under continuing development. However, first
calculations in the framework of this model indicate that in
provide some novel predictions for jet observables and for
two-particle (particulary for jet-like) azimuthal correlations in
general.

\section*{Acknowledgments}
Work partially supported by project (RNP.2.2.2.3.6039) of the
programme "Development of the scientific potential of the highest
school (2006-2008 years)" of the Federal Agency of Education of
the Russian Federation.


\begin{thebibliography}{6}

\bibitem{Adams-2004}J. Adams {\it et al.}, \Journal{Phys.Rev. Lett.}{93}{252301}{2004}.

\bibitem{Adams-2005}J. Adams {\it et al.}, \Journal{Nucl.Phys.}{A757}
{102}{2005}.

\bibitem{Adler-2003}C. Adler {\it et al.}, \Journal{Phys.Rev.Lett.}{90}
{032301}{2003}.

\bibitem{Bielcikova-2004} J. Bielcikova {\it et al.}, \Journal{Phys. Rev.}{C69}{021901(R)}{2004}.

\bibitem{Wang-2004}F. Wang, \Journal{J.Phys.}{G30}{1299}{2004}.

\bibitem{Dremin-2006}I.M. Dremin, \Journal{Nucl.Phys.}{A767}{233}{2006}.

\end{thebibliography}
\end{document}